\documentclass[
preprintnumbers,aps,pra,longbibliography,reprint,a4paper,nofootinbib,floatfix,superscriptaddress
]{revtex4-2}
\usepackage[utf8]{inputenc}
\usepackage[english]{babel}
\usepackage[T1]{fontenc}

\usepackage{amsmath}
\usepackage{amssymb}
\usepackage{braket}
\usepackage{units}
\usepackage{mathtools}
\usepackage{graphicx}
\usepackage{hyperref}

\hypersetup{citecolor=blue}
\hypersetup{colorlinks=true}
\hypersetup{linkcolor=blue}
\hypersetup{urlcolor=blue}

\DeclareMathOperator\tr{tr}
\DeclarePairedDelimiter\abs\lvert\rvert
\DeclarePairedDelimiter\norm\lVert\rVert
\newcommand*\dd{\mathrm{d}}
\newcommand*\ee{\mathrm{e}}
\newcommand*\ketbra[2]{\ket{#1}\!\bra{#2}}
\newcommand*\ketbras[3]{\ket{#2}_{#1}\!\bra{#3}}
\newcommand*\nn{\hat{\mathrm{U}}}
\newcommand*\n{\hat{\mathrm{u}}}
\newcommand*\proj[1]{\ketbra{#1}{#1}}
\newcommand*\projs[2]{\ketbras{#1}{#2}{#2}}
\newcommand\PT{\Lambda}
\renewcommand*\Im{\operatorname{Im}}
\renewcommand*\Re{\operatorname{Re}}
\renewcommand*\rho\varrho
\renewcommand*\vec\boldsymbol

\begin{document}

\preprint{SI-HEP-2026-10}
\preprint{P3H-26-031}

\author{Hai-Chau Nguyen}
\email{chau.nguyen@uni-siegen.de}
\affiliation{Naturwissenschaftlich-Technische Fakultät, Universität Siegen, Walter-Flex-Straße 3, 57068 Siegen, Germany}

\author{Gilberto Tetlamatzi-Xolocotzi}
\email{gtx@physik.uni-siegen.de}
\affiliation{Naturwissenschaftlich-Technische Fakultät, Universität Siegen, Walter-Flex-Straße 3, 57068 Siegen, Germany}

\author{Carmen~Diez~Pardos}
\email{carmen.dpardos@uni-siegen.de}
\affiliation{Naturwissenschaftlich-Technische Fakultät, Universität Siegen, Walter-Flex-Straße 3, 57068 Siegen, Germany}

\author{Otfried Gühne}
\email{otfried.guehne@uni-siegen.de}
\affiliation{Naturwissenschaftlich-Technische Fakultät, Universität Siegen, Walter-Flex-Straße 3, 57068 Siegen, Germany}

\author{Matthias Kleinmann}
\email{matthias.kleinmann@uni-muenster.de}
\affiliation{Department for Quantum Technology, Universität Münster, Heisenbergstraße 11, 48149 Münster, Germany}
\affiliation{Naturwissenschaftlich-Technische Fakultät, Universität Siegen, Walter-Flex-Straße 3, 57068 Siegen, Germany}

\title{Optimised Inference of Quantum Phenomena in High-Energy Collider Experiments}

\begin{abstract}
Entanglement, a fundamental phenomenon of quantum theory, has recently been observed in processes in high-energy physics. This opens new avenues for probing quantum effects in relativistic regimes, but also poses conceptual and technical challenges. We develop a general framework based on shadow tomography techniques for characterising spin-spin correlations in collider experiments. This improves the analysis of spin-spin entanglement, where relativistic motion couples spin and momentum and the momenta of the investigated particles are not under experimental control. As a proof of concept we illustrate the application of our formalism to top quark pair production at the Large Hadron Collider at CERN. The framework, however, is general and flexible and can be readily applied to more complex final states and systems with more particles.
\end{abstract}

\maketitle

Entanglement is a fundamental feature of quantum mechanics, playing a central role in our understanding of the theory and underpinning the development of modern quantum technologies. Nowadays, quantum entanglement as well as other quantum phenomena are routinely demonstrated across a wide range of experiments, for example using photons \cite{Aspect1982a, Gregor1998a}, trapped ions \cite{Rowe2001a, Haeffner2005a}, arrays of neutral atoms \cite{bluvstein2022a}, and superconducting qubits \cite{Cao2023a}. These experiments entangle quantum systems on purpose and the entanglement is a key property towards enabling future quantum experiments and quantum technologies. In addition, entanglement has been verified recently in high-energy collider experiments, more specifically by the ATLAS and CMS collaborations \cite{ATLAS:2023fsd, CMS:2024pts, CMS-TOP-23-007, ATLAS:2026hye}. In these experiments, collision processes generate particle pairs with entangled spin degrees of freedom, observing
the entanglement in a natural process. Probing entanglement in these systems allows one to verify theoretical models of the underlying processes involved \cite{Cheng:2024btk}. Still, there are at least two fundamental obstacles towards a detailed analysis and a full understanding of quantum correlations in high-energy experiments, given the nature of the scattering experiments.

First, the spins and momenta of relativistic particles are in general coupled under reference frame transformations \cite{sternberggroup1995}. Consequently, ignoring the momenta of the particles and concentrating on the spin state leads to a state with properties that are not Lorentz covariant, in particular, whether a state is entangled or not can depend on the reference frame \cite{peresquantum2004}. It would be advantageous to keep the momenta of the particles in the description of the spin state, resulting in an ensemble of spin density operators parametrised by momenta. But this is inherently difficult due to the nature of scattering experiments: The momenta of the parent particles are measured along with the spin and spread out over a wide range of values. A key question is how to analyse tomographic data if only few data points for narrow ranges of the momenta are given.

Second, in collider experiments the direction of the final-state decay products is used to infer the spin state of the parent 
particles \cite{boudjemamodel2009, goncalvesquantum2025}. The measurement of spin observables relies on the functional dependence of the normalised differential cross-section $ \frac{1}{\sigma} \frac{\dd\sigma}{\dd\Omega} $ on the spin state of the parent particles. Technically, the extraction of the spin information from the direction $(\vartheta,\varphi)=\Omega$ of the decay products is described via a spin operator $F(\Omega)$, and the normalised differential cross-section relates to the two-spin state $\rho$ as
\begin{equation}
    \frac1\sigma \frac{\dd\sigma}{\dd\Omega_1\dd \Omega_2} = \tr[\rho\, F(\Omega_1)\otimes F(\Omega_2)].
\end{equation}
In practice, the measurement model $F(\Omega_1)\otimes F(\Omega_2)$ is assumed to be known and fixed. Testing and verifying this assumption is essential, for instance in a thorough search for new physics in collider data with minimal assumptions.

In this paper we tackle both these obstacles by generalising shadow tomography methods \cite{huangpredicting2020, nguyenoptimizing2022}, originally developed in quantum information theory, to the analysis of high-energy collider experiments. First, for the spin-momentum coupling problem the technique of classical shadows provides the means to evaluate momentum-dependent spin observables directly, without reconstructing the momentum-resolved spin states. Second, we demonstrate that using shadow tomography, experimental data can also be used to verify the measurement model independently of the spin state, thus granting access to observables that can test novel aspects of our understanding of collider experiments.

To illustrate our approach, we perform an analysis of spin correlations in top quark pair ($t\bar t$) production in proton-proton collisions using Monte Carlo simulations. In particular, top-antitop quark pairs are simulated considering collisions at a centre-of-mass energy ($\sqrt{s}$) of \unit[13]{TeV}, which corresponds to the energy of the collisions at the Large Hadron Collider (LHC) during 2015--2018 and hence the datasets used in Refs.~\cite{ATLAS:2023fsd, CMS:2024pts, CMS-TOP-23-007}. We then illustrate how classical shadows can be constructed from the data, and how they can be used to analyse momentum-dependent entanglement of the produced $t\bar t$ as well as to provide consistency checks of the measurement model.

\emph{Momentum-spin ensemble of scattered particles.}---%
We consider the products of a scattering process where each scattering event involves $n$ relativistic particles with spins $(s_1,s_2,\dotsc,s_n)$. A measurement of the 4-momenta yields momenta $P=(p_1, p_2,\dotsc,p_n)$ distributed according to a probability distribution $h(P)\dd^{4n}\!P$. For each scattering event, given the observed momenta $P$, one then has a condition $n$-spin density operator $\rho_P$ and the expectation value of an $n$-spin observable $A$ conditioned on $P$ is $\braket{A}_P = \tr(\rho_PA)$. However, measuring this expectation value as a mean of observations is not possible because each scattering event yields different momenta $P$ and only a single outcome for $A$, namely one of the eigenvalues of $A$. Therefore, one considers the momentum-averaged expectation value,
\begin{equation}\label{eq:pmean}
    \braket{A} = \int \dd^{4n}\!P\, h(P) \tr[\rho_P A(P)],
\end{equation}
where the observable $A$ can depend on $P$.

Due to the relativistic nature of high-energy scattering events, one needs to discuss the effect of a global Lorentz transformation $\Lambda$ on spin observables. Crucially, when a particle with momentum $p$ is transformed by $\Lambda$, the spin density operator undergoes a Wigner rotation. We denote the Lorentz boost of a particle at rest to momentum $p$ by $L(p)$. The associated Wigner rotation is then given by $R(\Lambda,p)= L^{-1}(\Lambda p) \Lambda L(p)$ and its unitary representation for spin $s$ is then $D_{s}[R(\Lambda, p)]$. The spin density operator of $n$ particles in the new reference frame reads $D(\Lambda,P) \rho_{\Lambda P} D(\Lambda,P)^\dag$ where $D(\Lambda, P)= D_{s_1}[R(\Lambda, p_1)]\otimes D_{s_2}[R(\Lambda, p_2)]\otimes\dotsm\otimes D_{s_n}[R(\Lambda, p_n)]$ \cite{sternberggroup1995, Scheck2013}. A momentum-averaged expectation value as in Eq.~\eqref{eq:pmean} is hence Lorentz invariant if $A(P)$ transforms to $D(\Lambda, P) A(\Lambda P)D(\Lambda, P)^\dag$. 

\emph{Spin measurements of decaying particles.}---%
The spin information of unstable particles such as top quarks is determined through the information carried by its final state decay products. For instance in the decay process $t\to b \ell^+ \nu_{\ell}$ the spin information of the top quark can be inferred by analysing the angular distribution of the final state lepton $\ell^+$. We assume here that the direction of only one of the decay products is used, yielding a unit-vector $\n=(\cos\varphi \sin\vartheta, \sin\varphi\sin\vartheta,\cos\vartheta)$ in the rest frame of the decaying particle, specifically in the helicity basis \cite{Baumgart2013}. From the quantum information perspective, the direction $\n$ presents an outcome of a generalised measurement \cite{Heinosaari:2012} of the spin of the decaying particle. The measurement is hence described by a positive operator-valued measure (POVM) $F(\n)$, that is, by a positive spin-$s$ operator $F(\n)$ with $\int \dd \n F(\n) = \openone$, where $\dd\n=\sin(\vartheta)\dd\vartheta\dd\varphi$ is the solid angle measure $\dd\Omega$. The probability distribution for observing the direction $\n$ is then $\tr[\rho F(\n)]\dd\n$. Here $\rho$ is the density operator describing the spin-$s$ state of the parent particle. Note that $\tr[\rho F(\n)]$ is the normalised differential cross-section $\frac1\sigma\frac{\dd\sigma}{\dd\Omega}$. Quite generally one can show that in the rest frame the rotational symmetry of the evolution of the system implies the covariant transformation of $F(\n)$ under rotation, namely, $D_s[R] F(\n) D_s[R] = F(R \n)$ for rotation matrix $R \in \mathrm{SO}(3)$. Further symmetry considerations then imply \cite{nguyen2020symmetries}
\begin{equation}\label{eq:effect}
    F(\n) = \frac{2s+1}{4\pi}\sum_{m=-s}^{+s} \alpha_m \ket{\n,m}\bra{\n,m},
\end{equation}
where $\ket{\n,m}$ denotes the spin-$s$ state in quantisation direction $\n$; see Appendix~\ref{app:spin-in-weak-decay}. The coefficients $\alpha_m$ are directly related to the spin analysing power \cite{Brandenburg2002} and have to be computed or estimated from independent experiments. Special cases of Eq.~\eqref{eq:effect} are known in different forms, see, for example, Refs.~\cite{boudjemamodel2009, goncalvesquantum2025}. For spin-$\frac12$ particles with spin state $\rho$, the normalised differential cross-section is
\begin{equation}
    \frac1\sigma \frac{\dd \sigma}{\dd\Omega}= \tr[\rho F(\n)]= \frac{1+(\alpha_{+\frac12}-\alpha_{-\frac12}) \vec B\cdot \n}{4\pi},
\end{equation}
where $\vec B$ is the Bloch vector of $\rho$, that is, $B_i=\tr(\rho\sigma_i)$. For a system of $n$ spins, the decay product for each particle is recorded, giving $\nn=(\n_1,\n_2,\dotsc,\n_n)$ and the corresponding measurement $F(\nn) = F_1(\n_1)\otimes F_2(\n_2)\otimes\dotsm\otimes F_n(\n_n)$.

\emph{Shadow tomography of momentum-spin ensembles.}---%
To estimate $\braket A$ in Eq.~\eqref{eq:pmean} from experimental data one faces the difficulty that the observations consist only of the measured momenta and decay directions, $(P,\nn)$, jointly distributed according to $h(P,\nn)=h(P)\tr[\rho_P F(\nn)]$. Since in the data the 4-momenta $P$ of the produced particles in every event are different, $\tr[\rho_PA(P)]$ is not accessible, hindering an estimation of $\braket A$. The basic idea to address this problem is to find a function $a(P,\nn)$ such that $\int \dd^n\nn\, F(\nn)a(P,\nn)=A(P)$, enabling
\begin{equation}\begin{split}
    \braket A
    &= \int \dd^{4n}\!P          \, h(P) \tr[\rho_PA(P)]\\
    &= \int \dd^{4n}\!P\,\dd^n\nn\, h(P) \tr[\rho_PF(\nn)] a(P,\nn)\\
    &= \int \dd^{4n}\!P\,\dd^n\nn\, h(P,\nn)a(P,\nn).
\end{split}\end{equation}
Here, $\dd^n\nn$ is a shorthand for $\dd \n_1 \dd \n_2\dotsm\dd \n_n$. Since $(P,\nn)$ is sampled according to the distribution $h(P,\nn)$, it is now a simple matter of computing the average $\bar A= \frac 1N\sum_i^N a(P_i,\nn_i)$ to get an estimate for $\braket A$. That is, when $N$ events $i=1,\dotsc, N$ and according observations $(P_i, \nn_i)$ have been collected, then $\bar A$ converges to $\braket A$ as $N\to\infty$ with variance $\Delta \bar A=\Delta a(P,\nn)/N$.

It remains to provide the function $a(P,\nn)$. To this end, we employ the recently developed concept of shadow tomography \cite{huangpredicting2020}. In a nutshell, shadow tomography amounts to convert each observed record $\nn$ into a single-event spin operator $T(\nn)$, called the classical shadow~\cite{huangpredicting2020}. For a single event, the classical shadow does not need to be a valid quantum state, but it yields back the spin state on average, that is, $\braket{T} = \int \dd^n\nn\tr[\rho F(\nn)]T(\nn) = \rho$ for any density operator $\rho$. Hence $a(P,\nn)=\tr[T(\nn)A(P)]$ satisfies $\int \dd^n\nn \,F(\nn) a(P,\nn)= A(P)$ by virtue of
\begin{multline}
    \tr\left[\rho\int \dd^n\nn F(\nn) a(P,\nn)\right]\\
    = \int \dd^n \nn \tr[\rho F(\nn)] \tr[T(\nn)A(P)]=\tr[\rho A(P)]
\end{multline}
holding for all $\rho$. In Appendix~\ref{app:constr-shadow} we show that for a spin-$s$ measurement of the form of Eq.~\eqref{eq:effect} one can use
\begin{equation}
    T(\n) = \sum_{m=-s}^{+s} \beta_m \proj{\n,m},
\end{equation}
where the coefficients $\beta_m$ are given in terms of the coefficients $\alpha_m$. For example, for $s=\frac12$ one finds $\beta_m=\frac12[1+3/(\alpha_m-\alpha_{-m})]$. Clearly, for $n$ spins we can use $T(\nn)= T_1(\n_1)\otimes T_2(\n_2)\otimes\dotsm \otimes T_n(\n_n)$.

\emph{Spin entanglement in relativistic momentum-spin ensembles.}---%
We illustrate the use of shadow tomography for the detection of spin entanglement. According to entanglement theory \cite{Guhne:2009PR, Horodecki:2009RMP}, the quantum state of two particles is separable, if its density operator is a stochastic mixture of product states, that is, if it admits a decomposition of the form
\begin{equation}
    \rho_\mathrm{sep} = \sum_{i} q_i \proj{\psi_i}\otimes \proj{\phi_i},
\end{equation}
where the $q_i$ form a probability distribution, $q_i>0$ 
with $\sum_i q_i =1$. Conversely, a quantum state is entangled if it is not separable. A spin state $\rho_P$ with fixed momenta $P$ has Lorentz-covariant entanglement properties \cite{Alsing-2002-LorentzInvariance} because the induced spin-rotation $D(\Lambda, P)= D_{s_1}[R(\Lambda, p_1)]\otimes \dotsm\otimes D_{s_n}[R(\Lambda, p_n)]$ is a product of local unitary transformations and entanglement does not change under such transformations. Hence the classification of $\rho_P$ as entangled is invariant under Lorentz transformations. However, as we argued above, neither $\rho_P$ nor expectation values like $\braket{A}_P= \tr(\rho_P A)$ are accessible in experiments and hence one cannot directly show that $\rho_P$ is entangled.

For non-relativistic particles, it is common practice to consider the averaged spin density operator $\int \dd^{4n}\!P\, h(P)\rho_P$ \cite{peresquantum2004}. This is well-justified if the momenta are experimentally inaccessible and if the momenta are non-relativistic. However, for collider experiments the momenta are often available and the involved momenta are relativistic. Crucially, averaging over the momenta creates difficulties in interpreting entanglement under Lorentz transformations, since, after averaging, a transformation of reference frames can turn an averaged density operator from entangled to separable and vice versa~\cite{peresquantum2004}.

This conflict can be addressed by using entanglement witnesses as in Ref.~\cite{ATLAS:2023fsd, CMS:2024pts, CMS-TOP-23-007}. An entanglement witness is an observable that has positive expectation value for all separable states. A negative expectation value of the witness thus indicates entanglement \cite{Guhne:2009PR, Horodecki:2009RMP}. To characterise the entanglement of the momenta-parametrised spin density operators $\rho_P$, we allow the entanglement witnesses $W$ to depend on the momenta $P$. Since $\braket{W}_P \ge 0$ must hold for all separable states at fixed momenta $P$, observing a momentum-average $\braket{W}<0$ implies that $\rho_P$ must be entangled, at least for some momenta $P$.

Such momentum-dependent witnesses $W(P)$ can be tailored based on a theoretical prediction of the density operator $\rho(P)$, for example by using the partial-transpose based witness \cite{Horodecki:2009RMP, Guhne:2009PR}
\begin{equation}\label{eq:optwit}
    W_{\PT}(P) = \PT\left[\,\proj{\phi(P)}\,\right],
\end{equation}
with $\ket{\phi(P)}$ the eigenvector corresponding to the minimal eigenvalue of $\PT[\rho(P)]$ and $\PT$ the partial transpose of the second spin, $\PT[\, \ketbra{ij}{kl}\,] = \ketbra{il}{kj}$. In Appendix~\ref{app:evaluation} we discuss the construction of this witness for the spin of top-antitop pairs generated in proton-proton collider experiments.

\emph{Data consistency investigation.}---%
The validity of the procedure to estimate a spin-observable $\braket A$ from the observation mean $\bar A$, as outlined above, depends crucially on the validity of the description of the spin measurement via Eq.~\eqref{eq:effect}. The form of $F(\n)$ is based on (i) symmetry assumptions, (ii) the spin $s$ of the particle, and (iii) on the coefficients $\alpha_m$. For example, if the measurement model assumes that the measurement contains more noise than the actual experiment, then the estimate for an observable $A$ can exceed its maximal eigenvalue. More general tests of the measurement model can be developed using methods in the spirit of Refs.~\cite{morodercertifying2013, Smania2025, Freund2026}: We first note that the the observations $(P_i,\nn_i)$ provide us access to momentum-average expectation values of the type
\begin{equation}
    \braket{v} = \int \dd^{4n}\!P\,\dd^n\nn\, h(P,\nn)v(P,\nn),
\end{equation}
via $\bar v=\frac 1N \sum_i v(P_i,\nn_i)$. In principle, $v$ can be any momentum-spin function. We discuss here two types of functions that enable strong tests on the validity of the description of the spin measurement.

For the first type one chooses $v(P,\nn)=\tr[T(\nn)V(P)]$ with $V(P)$ a positive semidefinite spin operator and the classical shadow $T(\nn)$. If the measurement model is valid, then it $\braket v\ge 0$ follows. Conversely, if $\bar v<0$ is observed, this is a strong indication to discard the measurement model. For example for spin $\frac12$ and a system with spin state is $\ket\psi$, one can choose the spin operator $V=\openone-\proj\psi$. If $\braket{v}<0$ then $\alpha_{\frac12}$ in the measurement model is closer to $\frac12$ than in the experiment and hence the inadequacy of the model can be detected from the experimental data.

For the second type one tests whether the data can be faithfully described by spin-operators with given spin $s$. For this we note that the spherical harmonics $Y_{l,m}$ vanish on average, $\braket{Y_{l,m}}=\int \dd\n \tr[\rho F(\n)]Y_{l,m}(\n)=0$, for all $m=-l,\dotsc,l$ and $l\ge 2s+1$ \cite{Varilly1989}. We now consider $v(P,\nn)$ to be a linear combination of such spherical harmonics, that is, $v(P,\nn)= \sum_{\vec m} v_{\vec m} Y_{l_1,m_1}(\n_1)Y_{l_2,m_2}(\n_2)\dotsm Y_{l_n,m_n}(\n_n)$ with at least one $l_k\ge2s_k+1$ in each summand, $s_k$ the spin of particle $k$, complex coefficients $v_{\vec m}$ and $\vec m=(m_1,m_2,\dotsc,m_n)$. Then, observing $\bar v<0$ or $\bar v>0$ may indicate a failure of the theoretical model or hint towards an inconsistency of the applied data evaluation methods. For example, for a spin-$\frac12$ particle, one can use $v(\n)=Y_{2,0}(\n)=\sqrt{5/16\pi} (3 \cos(\vartheta)^2-1)$.

\begin{figure}
    \centering
    \includegraphics[width=\linewidth]{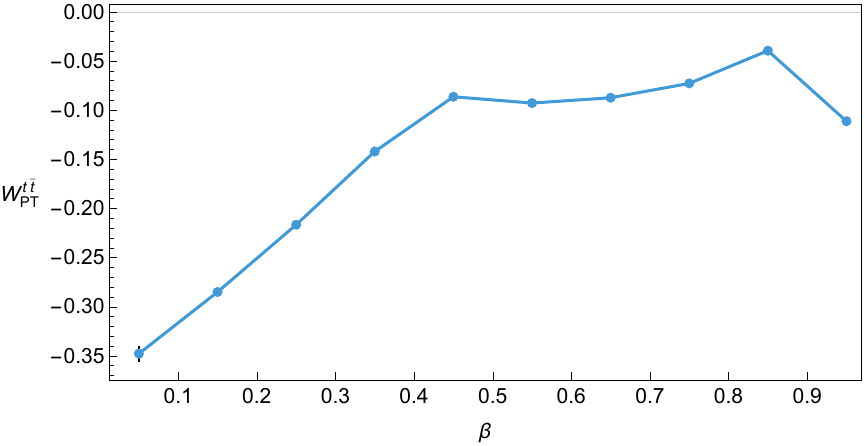}
    \caption{Average values of partial-transpose based witness $W_{\PT}^{t\bar t}(P)$ for the $t\bar t$ spin state estimated from Monte Carlo simulated events. The witness is constructed according to Eq.~\eqref{eq:optwit} using the spin density operator $\rho(P)$ provided in Ref.~\cite{afikentanglement2021}, see Appendix~\ref{app:evaluation}. The simulated data are shown in the bins $k \le\beta < k+0.1$ for $k=0.0,0.1,\dotsc,0.9$ with $\beta = \abs{\vec p_t}/E_t$ the velocity of the top quark in the centre of mass frame of $t\bar t$. Only events are selected where the top quark momentum $p_t$ is such that the predicted state $\rho(P)$ is entangled. This enables the detection of entanglement for all $\beta$. For each bin we show the mean value of $W^{t\bar t}_{PT}$ together with the statistical uncertainty.
    }
    \label{fig:witness}
\end{figure}

\begin{figure}
    \centering
    \includegraphics[width=\linewidth]{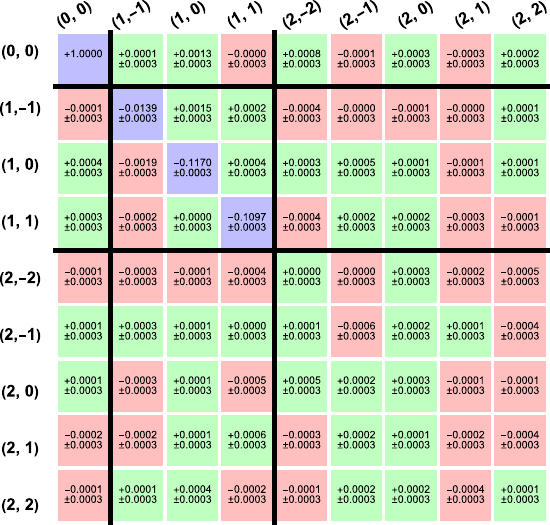}
    \caption{Correlation matrix of the $t\bar t$ spin obtained using $10^7$ Monte Carlo simulated events. The uncertainties quoted correspond to the statistical uncertainty, which is known to underestimate the uncertainty in experiments. Every entry, except for the upper left 4\texttimes 4 matrix, is predicted to be zero. The correlations are computed as the mean value of $4\pi Y^{\Re}_{l,m}(\n_t)Y^{\Re}_{l',m'}(\n_{\bar t})$, where $Y^{\Re}$ denotes the real spherical harmonics, see Appendix~\ref{app:evaluation}. In the matrix, $(l,m)=(0,0),(1,-1),(1,0),(1,+1),\dotsc(2,+2)$ indexes the rows and $(l',m')$ the columns; the singlet state $\ket{\psi^-}$ would yield a diagonal matrix with entries $1,-\frac13,-\frac13,-\frac13,0,\dotsc,0$. Only the blue entries are expected to be nonzero; positive and negative entries are green and red, respectively.
    }
    \label{fig:correlations}
\end{figure}

\emph{Proof of concept study with Monte Carlo simulation.}---%
To illustrate the practical application of our method, we focus on $t\bar t$ production in proton-proton collisions $p p \to t \bar t$. This process was considered recently at the LHC \cite{ATLAS:2023fsd, CMS:2024pts, CMS-TOP-23-007} verifying entanglement in $t\bar t$. The events were produced using the \textsc{Powheg~Box} Monte Carlo generator \cite{Nason:2004rx, Frixione:2007vw, Alioli:2010xd}. We employed the \texttt{hvq} module for heavy-quark production \cite{Frixione:2007nw}. Thus, we generated $10$ million $p p \to t \bar t$ dileptonic events at $\sqrt{s} = \unit[13]{TeV}$ at next-to-leading order (NLO) accuracy in quantum chromodynamics (QCD). For the purposes of this proof-of-concept study, only parton-level information was used. We leave the hadronisation effects for future investigations.

The details of the event generation are as follows. The NNPDF3.1 NLO \cite{Ball2017} parton distribution functions were used via the LHAPDF interface. The renormalisation and factorisation scales were set to their default dynamical values. A five-flavour scheme was adopted for the QCD calculation. The top quark mass was fixed to $m_t = \unit[172.5]{GeV}$ and we considered the corresponding decay width $\Gamma_t = \unit[1.32]{GeV}$, while the W-boson mass and width were set to $m_W = \unit[80.399]{GeV}$ and to $\Gamma_W = \unit[2.085]{GeV}$, respectively. The remaining fermion masses were set to their standard values. Radiation from the hardest emission was regulated using the \textsc{POWHEG} damping parameter, set to $h_{\mathrm{damp}} = \unit[258.75]{GeV}$. In addition, a minimal transverse-momentum cut of $k_T^{\mathrm{min}} = \unit[5]{GeV}$ was applied at the Born level. The $t\bar t$ pairs were decayed within \textsc{POWHEG} itself, restricting the final state to dileptonic channels involving electrons and muons only. Thus we considered $t\rightarrow b \ell^+ \nu_{\ell}$ and correspondingly for the $\bar{t}$. Finally, the event generation was performed using standard \textsc{POWHEG} integration settings. To analyse the data we use the analysis framework \textsc{HWSim} \cite{hwsim} addon to \textsc{HERWIG 7} \cite{Bahr:2008pv, Bellm:2017bvx, Gieseke:2011na, Arnold:2012fq, Bellm:2013hwb,Bellm:2015jjp}.

This setup is enough to demonstrate that (i) shadow tomography methods can be used to evaluate a momentum-dependent entanglement witness allowing to detect entanglement across the full energy range, see Fig.~\ref{fig:witness} and Appendix~\ref{app:evaluation}, and (ii) the data consistency conditions can be used to test the measurement model $F(\nn)$, see Fig.~\ref{fig:correlations} and Appendix~\ref{app:evaluation}, where also an example for a failure of the conditions is discussed.

\emph{Conclusion.}---%
We have pointed out that using the language of generalised measurements and classical shadows provides a suitable framework to analyse spin in collider experiments, enabling the full use of experimental data, a covariant understanding of spin entanglement, and novel consistency test on experimental data. As an illustration, we applied this framework to simulated parton-level data for top-antitop quark production in proton-proton collisions. We demonstrated the detection of entanglement for all velocities using a momentum-dependent entanglement witness and we showed that the measurement model---encoded in the normalised differential cross-section---can be tested from the same data. As expected, no discrepancies were found in the simulated data, but applying these tests to experimental data could reveal unexpected deviations. Our methods are general and can be applied to more complex scenarios in particular with more general final states $\rho(P)$ and more than two particles. Moreover, the language of generalised measurements is sufficiently flexible to incorporate further details of the physics at detector level, providing an important direction for further investigation with direct applications on real experimental data.

\emph{Acknowledgements.}--%
We thank Andreas Papaefstathiou for useful discussions on the Monte Carlo generators and his help with \textsc{HWSim}. We thank Tao Han for interesting physics discussions during his visits to Siegen. We also thank Tobias Huber for his valuable input in the early stages of the project. The computer simulations of this work were partially performed on the HPC cluster OMNI at Siegen University. This work was supported by 
the Deutsche Forschungsgemeinschaft (DFG, German Research Foundation), project numbers 563437167 and grant 396021762 -- TRR~257 ``Particle Physics Phenomenology after the Higgs Discovery'',
the project EINQuantum NRW,
the Sino-German Center for Research Promotion (Project M-0294), and 
the German Federal Ministry of Research, Technology and Space (Project BeRyQC, Grant No. 13N17292),
European Union’s Horizon 2020 research and innovation programme under the Marie Sklodowska-Curie grant agreement No 945422.

\bibliography{quark}

\appendix
\onecolumngrid
\clearpage
\fontsize{11}{14}\selectfont

\section{Spin measurement by momentum measurement of decay products}
\label{app:spin-in-weak-decay}

In experiments with fast-decaying particles, the spin degree of freedom is normally not directly accessible. Fortunately, decay products can reveal information about the spin state of the parent particle. The relevant case for this study is where the direction of the momentum of one of the decay products is directly related to the spin of the decaying particle.

To understand this from a general perspective, we consider a particle of spin $s$ at rest and in spin state $\ket{\psi}_S$. The particle then decays in several products. We assume that the momentum $p$ of at least one of the decay particles is tracked. We denote the direction of this momentum by $\n = \vec{p}/\abs{\vec{p}}$ and the corresponding degree of freedom by $T$. All other degrees of freedom are denoted by $Q$ and collected into the quantum number $\chi$, including, for example, the magnitude of the momentum $\abs{\vec{p}}$, the spin of the tracked particle, and all other decay products. The initial state of the system is therefore $\ket{i} = \ket{\phi}_{QT} \ket{\psi}_S$ and the final state is $\ket{f} = \ket{\chi,\n}_{QT} \ket{\mathrm{vac}}_S$, where $\ket{\phi}_{QT}$ is the joint initial state of the system $QT$ and $\ket{\mathrm{vac}}_S$ is the vacuum for the spin. The amplitude of transition is then given by $\braket{f|U|i}$ where $U$ is the asymptotic time evolution of the system. The probability density for the direction $\n$ is now given by
\begin{equation}
    P_{f|i}(\n) = \int\dd\chi \abs{\braket{f|U|i}}^2
    = \int \dd\chi \bra{\phi}_{QT} \bra{\psi}_S U^{\dag}
    \left( \projs{QT}{\chi,\n} \otimes \projs{S}{\mathrm{vac}} \right) U \ket{\phi}_{QT} \ket{\psi}_S.
\end{equation}
This can be rewritten as $P_{f|i}(\n) = \bra{\psi}_S \tilde F(\n) \ket{\psi}_S$. Then $\tilde F(\n)$ is the spin operator
\begin{equation}
    \tilde F(\n) = \bra\phi_{QT} U^\dag \left(\Pi(\n) \otimes \projs{S}{\mathrm{vac}} \right) U \ket\phi_{QT}
\end{equation}
with $\Pi(\n)=\int \dd \chi \projs{QT}{\chi,\n}$. Here we used the shorthand notation
\begin{equation}
\bra\phi_{QT} X_{QTS} \ket\phi_{QT} = \sum_{m,n=-s}^s\ketbras{S}{m}{n}\, \bra\phi_{QT}\bra{m}_S X_{QTS} \ket\phi_{QT}\ket{n}_S.
\end{equation}
Up to a normalisation factor, $\tilde F(\n)$ coincides with the positive operator-valued measure $F(\n)$ characterising the measurement of the spin via the direction $\n$ of the momentum of the decay product as defined in the main text (see also below).

We show now that, quite generality, if the evolution operator $U$ is rotational invariant, then $\tilde F(\n)$ is covariant under rotations. Let $R \in \mathrm{SO}(3)$ be a proper rotation. We use $R_{QT}$ and $R_S$ to denote its representation in the respective subsystems and $R_{QTS}=R_{QT}\otimes R_S$. (In order to ease the notation, we use here $R_X$ in place of the notation $D_X[R]$ in the main text.) Now we see that
\begin{equation}\label{eqap:covariance}
\begin{split}
    R_S \tilde F(\n) R_S^\dag &=
    \bra\phi_{QT} \left(R_{QT}\otimes R_S\right)
    \left(U^\dag \left(\Pi(\n) \otimes \projs{S}{\mathrm{vac}} \right) U \right)
    \left(R_{QT}\otimes R_S\right)^\dag \ket\phi_{QT}\\
    &=
    \bra\phi_{QT} U^\dag
    \left(R_{QT}\Pi(\n)R^\dag_{QT} \otimes \projs{S}{\mathrm{vac}}\right) U \ket\phi_{QT}\\
    &=
    \bra\phi_{QT} U^\dag\left( \Pi(R\n) \otimes \projs{S}{\mathrm{vac}} \right)U \ket\phi_{QT}
    =\tilde F(R\n).
\end{split}\end{equation}
In other words, $\tilde F(\n)$ transforms covariantly under rotation as a consequence of the rotational invariance of the evolution of the system. For obtaining this result, we have inserted the rotation $R_{QT}$ using the fact that $\projs{QT}\phi$ is invariant  under $R_{QT}$ (recall that the particle is at rest), and similarly, in the second step we used that $\projs{S}{\mathrm{vac}}$ is invariant under $R_S$. Furthermore, in the second step we used that the evolution operator $U$ commutes with the rotation $R_{QTS}$ of the whole system. In the final step, we used $R_{QT}\Pi(\n)R_{QT}^\dag = \Pi(R\n)$. In order to justify this, we first assume that the direction $\ket\n$ is an independent degree of freedom. More precisely, $Q$ and $T$ are independent and therefore $\Pi(\n)= \openone_Q \otimes \projs{T}\n$. Then $R_{QT}=R_Q\otimes R_T$ and $R_T\ket\n=\ket{R\n}$ implies the claim.

To derive the explicit form of $\tilde F(\n)$ as imposed by the covariance property, we follow the argument used in Ref.~\cite{nguyen2020symmetries}. Fixing a direction $\n$, we consider all rotations $R$ that leave $\n$ invariant, $R\n =\n$. From Eq.~\eqref{eqap:covariance} we see that $\tilde F(\n)$ commutes with all corresponding unitaries $R_S$. Because all these transformations are generated by the spin operator $S(\n)= \n\cdot \vec S$ with $\vec S=(S_x,S_y,S_z)$, the operator $\tilde F(\n)$ must commute with $S(\n)$ and hence it must be a linear combination of the spin projections onto direction $\n$ \cite{nguyen2020symmetries}, that is,
\begin{equation}\label{eqap:tildeeffect}
    \tilde F(\n) = \frac{2s+1}{4\pi} \sum_{m=-s}^{+s} \tilde{\alpha}_m \proj{\n,m}
\end{equation}
with $\ket{\n,m}$ the eigenvector to the eigenvalue $m$ of the spin operator $S(\n)$. Using the convention $\int \dd\n=4\pi$, we have $\int \dd \n \tilde{F} (\n)= \sum_m \tilde{\alpha}_m \openone$. The sum $\sum_m \tilde{\alpha}_m$ is nothing but the probability for the considered decay channel to actually take place. Post-selecting only events where the decay channel actually takes place, one obtains a (normalised) positive operator-valued measure $F(\n)= \tilde{F}(\n)/(\sum_m \tilde{\alpha}_m)$. In other words, $F(\n)$ follows the same expansion~\eqref{eqap:tildeeffect} with coefficients normalised as $\sum_{m} \alpha_m=1$.

\section{Construction of the classical shadow}\label{app:constr-shadow}
We consider a measurement where the outcome is a direction $\n$ on the 2-sphere, described by the positive operator-valued measure (POVM) $F(\n)\dd\n$ such that the probability distribution for observing $\n$ for a system described by the density operator $\rho$ is given by
\begin{equation}
    p(\rho;\n)\dd\n= \tr(\rho F(\n))\dd\n.
\end{equation}
Of particular interest for us are covariant POVMs of a spin-$s$ system. Those are of the form, see Appendix~\ref{app:spin-in-weak-decay},
\begin{equation}
    F(\n) = \frac{2s+1}{4\pi}\sum_{m=-s}^s \alpha_m\proj{\n,m},
\end{equation}
where $\alpha_m$ are nonnegative numbers with $\sum_m \alpha_m=1$. In this equation we assumed $\int \dd\n = 4\pi$ and wrote $\ket{\n,m}$ for the eigenvector to the eigenvalue $\hbar m$ of $\n\cdot \vec S$, where $\vec S$ are the three spin operators for spin $s$, With azimuthal and polar angles $\varphi$ and $\vartheta$, respectively, that is, $\n=(\cos(\varphi)\sin(\vartheta),\sin(\varphi)\sin(\vartheta),\cos(\vartheta))$, one has $\dd \n = \sin\vartheta\dd\vartheta\dd\varphi$ and
\begin{equation}
    \ket{\n,m}
    = \ee^{-i\varphi S_3/\hbar}\ee^{-i\vartheta S_2/\hbar}\ket{m}
    = \sum_{m'=-s}^sD^s_{m',m}(\varphi,\vartheta,0)\ket{m'},
\end{equation}
where $D^s$ denotes the Wigner $D$-matrix for spin $s$ and $\ket m=\ket{(0,0,1),m}$.

For shadow tomography, the shadow operator $T(\n)$ is such that when we sample from $p(\rho;\n)\dd\n$ then we obtain back the state $\rho$ as mean value of $T(\n)$, that is,
\begin{equation}\label{eqap:shadow}
    \rho=\braket{T}_\rho = \int \dd\n\, T(\n)\tr(F(\n)\rho).
\end{equation}
We mention that $T$ is not uniquely determined by this condition, unless the shadow is required to be covariant under rotations, $R_ST(\n)R_S^\dag = T(R\n)$. For an explicit construction one proceeds as follows. First, with the same argument as in Appendix~\ref{app:spin-in-weak-decay}, covariance yields the explicit form~\cite{nguyenoptimizing2022}
\begin{equation}
    T(\n)=\sum_m \beta_m \proj{\n,m}
\end{equation}
with $\sum_m\beta_m= \tr\braket{T}_\rho= \tr\rho=1$. Second, the coefficients $\beta_m$ can be determined by using that
\begin{equation}
    \int \dd\n \, T(\n)\otimes F(\n) = \sum_{m,m'}\ketbra m{m'}\otimes \ketbra{m'}m,
\end{equation}
is equivalent to Eq.~\eqref{eqap:shadow} to hold for all $\rho$. Then, by defining the operators
\begin{equation}
    L_{m,m'} =\frac{2s+1}{4\pi} \int \dd \n \ketbra{\n,m}{\n,m'}\otimes \ketbra{\n,m'}{\n,m},
\end{equation}
we arrive at the linear equation
\begin{equation}
    \sum_{m,m'} \beta_{m}\alpha_{m'} L_{m,m'} = \openone,
\end{equation}
which can be readily solved for $\beta_{m}$. Note, that $L_{m,m'}$ can be computed directly using the product rules for Wigner $D$-matrices. For example, for spin-$\frac12$ systems we obtain
\begin{equation}
    \beta_{\pm\frac12} = \frac12\left(1\pm\frac{3}{\alpha_{+\frac12}-\alpha_{-\frac12}}\right)
\end{equation}
and for spin-1 systems we find
\begin{equation}
    \beta_{\pm1}=\frac{1-\beta_0}2 \pm\frac1{\alpha_{+1}-\alpha_{-1}}\quad\text{and}\quad
    \beta_0 = \frac13\left(1-\frac{10}{\alpha_{+1}-2\alpha_0+\alpha_{-1}}\right).
\end{equation}

\section{Evaluation of the Monte Carlo simulations}
\label{app:evaluation}
\subsection{Transformation to the helicity reference frame}
In the Monte Carlo data, for each event the 4-momentum $p_t$ ($p_{\bar t}$) of the top (antitop) quark and the 4-momenta $p_a$, $p_b$ of the corresponding decay leptons are given in the lab frame with the beam axis $b=(1,0,0,1)$ in $z$-direction. We boost all momenta and the beam axis to the centre of mass system with velocity $\vec \beta_\mathrm{cms} = (\vec p_t+\vec p_{\bar t})/(p_t^0+p_{\bar t}^0)$, yielding $p'_t$, $p'_{\bar t}$, $p'_a$, $p'_b$, and $b'$. From here we boost $p'_a$ to the rest frame of the top with velocity $\vec \beta_t=\vec p'_t/p^{\prime0}_t$, and similarly we boost $p'_b$ to the rest frame of the antitop with velocity $\vec \beta_{\bar t}=\vec p'_{\bar t}/p^{\prime0}_{\bar t}=-\vec \beta_t$. Then we rotate the resulting vectors to the helicity frame, yielding $p''_a$ and $p''_b$, that is, we perform a rotation such that $\vec \beta_t$ becomes the $z$-direction, $\vec b'\times \vec\beta_t$ the $x$-direction, and $\vec \beta_t\times (\vec b'\times\vec\beta_t)$ the $y$-direction. Note that both lepton momenta are rotated in the same way. Finally, we flip the spacial components of $p''_b$, putting it into a left-handed basis. This flip ensures that the probability distribution over the normalised directions $\nn_a$ and $\nn_b$ is given by
\begin{equation}
    p(\rho;\nn_a,\nn_b)\dd \nn_a\dd\nn_b = \tr(\rho F[\nn_a]\otimes F[\nn_b])\dd \nn_a\dd\nn_b
\end{equation}
with $\rho$ the $t\bar t$ spin density operator in the helicity basis.

\subsection{Computing the correlation matrix}

To avoid complex numbers in the correlation matrix we use the real spherical harmonics
\begin{equation}\begin{split}
Y^{\Re}_{l,m}&=\sqrt 2(-1)^m \Im(Y_{l,-m}) \quad\text{for } m=-l,\dotsc,-1,\\
Y^{\Re}_{l,0}&= Y_{l,0},\\
Y^{\Re}_{l,m}&=\sqrt 2(-1)^m \Re(Y_{l,m}) \quad\text{for } m=1,\dotsc, l,
\end{split}\end{equation}
which also from an orthonormal function system. In terms of $x=\cos\varphi \sin\vartheta$, $y=\sin\varphi\sin\vartheta$, $z=\cos\vartheta$, the lowest order real spherical harmonics are
\begin{gather}
    Y^{\Re}_{0,0} = c_0\\
    Y^{\Re}_{1,-1} = c_1 y,\quad
    Y^{\Re}_{1,0} = c_1 z,\quad
    Y^{\Re}_{1,+1}=c_1 x\\
    Y^{\Re}_{2,-2} = c_2 xy,\quad
    Y^{\Re}_{2,-1} = c_2 yz,\quad
    Y^{\Re}_{2,0} = \frac{c_2}{\sqrt{12}}(3 z^2-1),\quad
    Y^{\Re}_{2,+1}= c_2 xz,\quad
    Y^{\Re}_{2,+2} = \frac{c_2}2 (x^2-y^2)
\end{gather}
with $c_0=\sqrt{1/4\pi}$, $c_1=\sqrt {3/4\pi}$, and $c_2=\sqrt{15/4\pi}$.

We briefly expand here on why for a spin-$s$ system, $\braket{Y_{l,m}}=0$ holds for $l\ge 2s+1$. We first note that
\begin{equation}\begin{split}
    \braket{\n,m|n}\!\braket{n'|\n,m} &= \sum_{l=0}^{2s}\sum_{k=-l}^l c_{m,n,n'}^{l,k} Y_{l,k}(\n)^*,
\end{split}\end{equation}
where the coefficients $c_{m,n,n'}^{l,k}$ can be computed using Clebsch--Gordan coefficients \cite{Varilly1989}. It follows that the probability density $p(\rho;\n)= \tr[\rho F(\n)] = \sum_m \alpha_m \braket{\n,m|\rho|\n,m}$ can be expressed as a linear combination of $Y_{l,k}(\n)^*$ with $l\le 2s$ and hence $\braket{Y_{l,m}}=\int \dd\n\,  p(\rho;\n)Y_{l,m}(\n)=0$  for $l\ge 2s+1$ due to the orthogonality of the spherical harmonics.

\subsection{Computing the variance in shadow tomography}
In the analysis in the main text we use the statistical uncertainty, computed as the sample standard deviation. Here we briefly point out that classical shadows allow us to efficiently compute the variance directly. For this we assume that the data $\n$ is produced by measuring the POVM $F(\n)$ and that we evaluate a function $g(\n)$ on the data. Hence,  given the state is $\rho$, we have the moments
\begin{equation}
    \braket{g^k} = \int  \dd \n \,g(\n)^k\tr[\rho F(\n)]
                    = \tr\left[\rho \int \dd \n\,g(\n)^kF(\n)\right]
                    = \tr(\rho G^{(k)})
\end{equation}
where we implicitly defined the spin operator $G^{(k)}$. While the value of $\braket{g^k}$ depends on the (unknown) state $\rho$, it can be efficiently bounded by computing the smallest and largest eigenvalue of $G^{(k)}$, due to $\lambda_{\min}(G^{(k)})\le \braket{G^k}\le \lambda_{\max}(G^{(k)})$. For example, for a spin-$\frac12$ particle and $g(\nn)=\sqrt{4\pi}Y^{\Re}_{l,m}(\n)$ with $l\ge 1$  one finds $G^{(2)}=\openone$ and $\tr(G^{(1)})=0$ and hence $(\Delta g)^2\le1$.

\subsection{Data consistency test}
To illustrate how the correlation matrix in  Fig.~\ref{fig:correlations} of the main text is able to detect physical inconsistencies, we generated 10 million $p\rightarrow t \bar{t}$ events in \mbox{MadGraph5\_aMC@NLO} \cite{Alwall:2011uj} at NLO accuracy in QCD, inducing decays into electrons and muons through \textsc{MadSpin}. This sample satisfies two consistency tests: (i) the mean value of the shadow operator is positive semidefinite, $\bar T\ge 0$ and (ii) this mean value is also close to the predicted state $\rho^\mathrm{LO}(p_t)$, having $\norm{\bar T-\braket{\rho^\mathrm{LO}}}<0.005$ at the production ratio $\eta=0.88$, see Eq.~\eqref{eqap:rholo} below. However, to the best of our knowledge this setup alone is not physically complete, for instance it does not lead to a fully consistent NLO calculation at partonic level unless extra gluon-emissions are introduced afterwards in a Monte Carlo generator like \textsc{Herwig 7}. The unphysical features lead to a correlation matrix  incompatible  with two decaying spin-$\frac12$ particles, since the corresponding evaluation fails to deliver zero entries everywhere outside the upper left 4\texttimes 4 matrix: for example we find $\bar v\ne 0$ at more than 10 standard deviations for the function $v(\nn)=4\pi Y_{2,0}(\n_t)Y_{2,0}(\n_{\bar t})$.

\subsection{Entanglement Witnesses}
To detect entanglement for all top quark momenta, we construct a momentum-dependent witness. In general, a density operator $\rho$ of two spin-$\frac12$ particles is entangled if and only if its partial transpose $\PT[\rho]$ has a negative eigenvalue \cite{Horodecki:2009RMP, Guhne:2009PR}. Here, the partial transpose is the transpose of either of the spins in a fixed basis; we use $\PT[\, \ketbra{ij}{kl}\,] = \ketbra{il}{kj}$. Given an entangled density operator $\rho$, we write $\ket\phi$ for the eigenvector of $\PT[\rho]$ with negative eigenvalue and let
\begin{equation}
  W_{\PT} = \PT[\, \proj{\phi} \,].
\end{equation}
By virtue of $\tr(A\PT[B])=\tr(\PT[A]B)$, this operator has negative expectation value for $\rho$ but positive expectation value for all separable states. Hence $W_{\PT}$ is an entanglement witness \cite{Horodecki:2009RMP, Guhne:2009PR}.

For the spin degrees of freedom in the $t\bar t$-system, we use the predicted spin states for the production of the $t\bar t$ pair via gluon fusion $\rho^{gg}$ and quark-antiquark interaction $\rho^{q\bar q}$; both are calculated in leading order in Ref.~\cite{afikentanglement2021}. The prediction is for the spin state in the helicity basis, and depends on the momentum $p_t$ of the top quark in the centre-of-mass reference frame of the $t\bar t$ system. We choose a fixed ratio of $84\%$ between the production channels, that is, we use the state
\begin{equation}\label{eqap:rholo}
    \rho^\mathrm{LO}(p_t) = \eta\, \rho^{gg}(p_t) + (1-\eta)\,\rho^{q\bar q}(p_t) \quad\text{with } \eta=0.84.
\end{equation}

One readily finds that the eigenvector corresponding to the negative eigenvalue of $\PT[\rho^{LO}(p_t)]$ is either of the Bell states $\ket{\Phi^+}=\frac1{\sqrt 2}(\ket{\uparrow\uparrow}+\ket{\downarrow\downarrow})$ or $\ket{\Psi^+}=\frac1{\sqrt 2}(\ket{\uparrow\downarrow}+\ket{\downarrow\uparrow})$. Accordingly, we define the witnesses $W_1=\PT[\,\proj{\Phi^+}\,]$ and $W_2=\PT[\,\proj{\Psi^+}\,]$, that is,
\begin{equation}\label{eqap:witness}
    W_{1,2}=\frac14\left(
       \sigma_0\otimes\sigma_0
       +   \sigma_1\otimes\sigma_1
       \pm \sigma_2\otimes\sigma_2
       \pm \sigma_3\otimes\sigma_3
       \right).
\end{equation}
These witnesses correspond to the entanglement markers $D$ and $\tilde D$ used, for example, in Refs.~\cite{ATLAS:2023fsd, CMS:2024pts}, via $D(\rho)=\tr[(4 W_1-\openone)\rho]/3$ and $\tilde D(\rho)=\tr[(4W_2-\openone)\rho]/3$.

\begin{figure}
    \centering
    \includegraphics[width=0.5\linewidth]{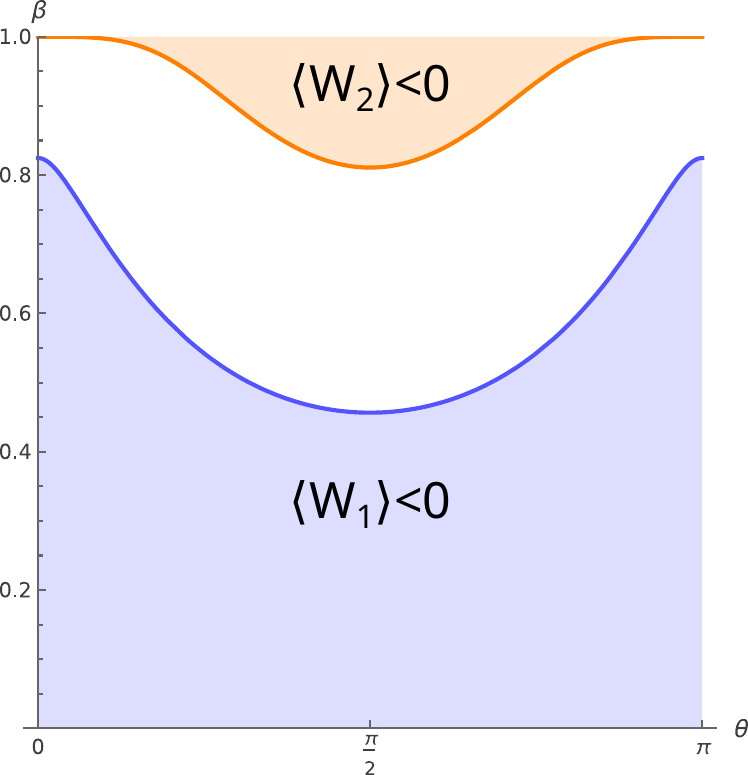}
    \caption{Regions where the entanglement witnesses $W_1$ and $W_2$, see Eq.~\eqref{eqap:witness}, have negative expectation value and hence detect entanglement. The expectation value is computed with respect to the spin state $\rho^\mathrm{LO}(p_t)$ in Eq.~\eqref{eqap:rholo}. Here, $\beta=\abs{\vec p_t}/p_{t,0}$ is the velocity of the top quark in the centre of mass system and $\theta=\arccos(\vec p_t\cdot \vec b/\abs{\vec p})$ the corresponding angle of the momentum with the beam pipe.}
    \label{figap:regions}
\end{figure}

\begin{figure}
    \centering
    \includegraphics[width=.6\linewidth]{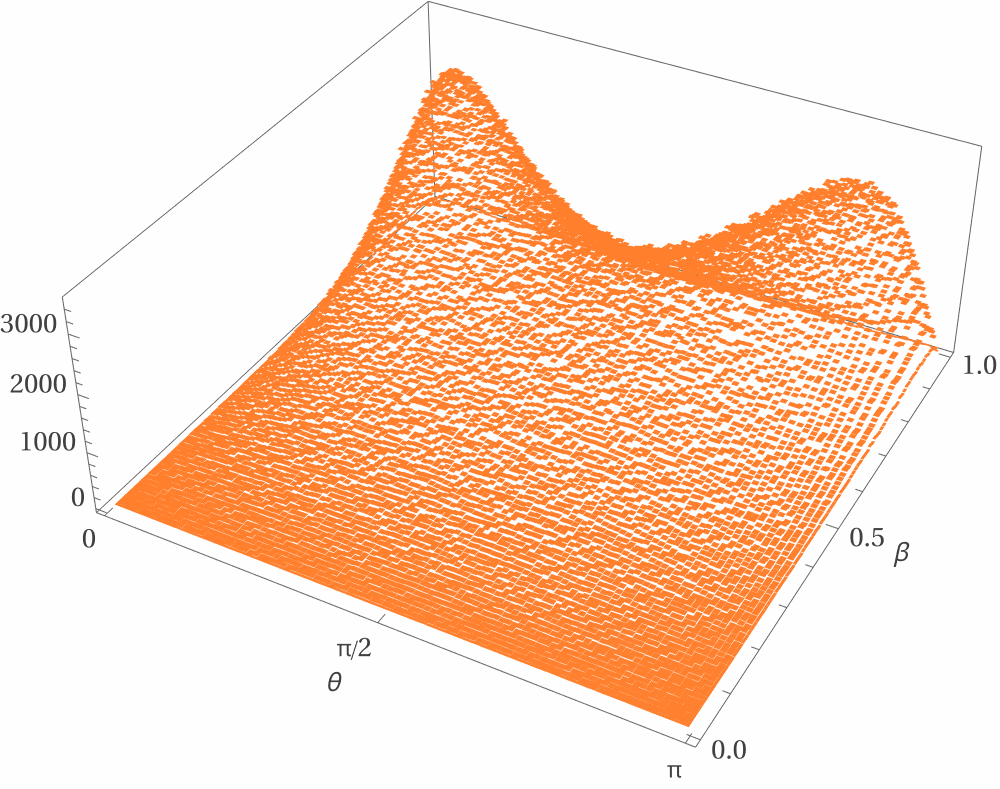}
    \caption{Histogram of the top quark momentum in the $t\bar t$ rest frame, obtained using $10^7$ Monte Carlo simulated events. $\beta$ is the velocity of the top quark, $\beta=\abs{\vec p_t}/p_{t,0}$, and $\vartheta$ is the angle of the top quark momentum with the beam pipe, $\theta=\arccos(\vec p_t\cdot \vec b/\abs{\vec p})$.
    }
    \label{figap:histogram}
\end{figure}

The witness $W^{t\bar t}_{\PT}(P)$ is now defined as $W_1$ if $\braket{W_1}_P=\tr[\rho^\mathrm{LO}(p_t) W_1]<0$ and as $W_2$ if $\braket{W_2}_P<0$. In Fig.~\ref{figap:regions} we show the regions where $\braket{W_1}$ or $\braket{W_2}$ are negative, parametrized by $\beta=\abs{\vec p_t}/p_{t,0}$ and $\theta=\arccos(\vec p_t\cdot \vec b/\abs{\vec p_t})$. Note that there is a region where the state is not entangled and hence $\braket{W_1}_P\ge 0$ and $\braket{W_2}_P\ge 0$. In the data analysis those events are filtered out:  It is valid to select the data based on the top quark momentum, because $p_t$ is independent of the spin information contained in the lepton momenta. Using the top quark momentum to remove events that are not expected to show entanglement can significantly improve the detection of entanglement, in particular because a large fraction of the produced $t\bar t$ pairs is produced at momenta where the predicted state is not entangled, see Fig.~\ref{figap:regions} and Fig~\ref{figap:histogram}.

We mention that the state $\rho^\mathrm{LO}(p_t)$ enters in the construction of the witness and determines how well the witness can detect entanglement in the experiment. However, the validity of the witness is independent of $\rho^\mathrm{LO}(p_t)$, that is, if a negative expectation value is observed, then this is always a proof of entanglement in the system.

\end{document}